\documentclass[]{aa} 

\usepackage{graphicx}
\usepackage{txfonts}

\usepackage{amsmath}	
\usepackage{amssymb}	
\usepackage{mathtools}

\usepackage{threeparttable}
\usepackage{longtable}
\usepackage{tabto}
\usepackage{adjustbox}
\usepackage{multicol} 
\usepackage{multirow}

\usepackage{color}
\usepackage{natbib,twoopt}
\usepackage[hyphenbreaks]{breakurl}
\usepackage[breaklinks]{hyperref}      
\bibpunct{(}{)}{;}{a}{}{,}             
\definecolor{cobalt}{rgb}{0.06, 0.2, 0.65}
\hypersetup{
  colorlinks,
  citecolor=blue,
  linkcolor=blue,
  urlcolor=blue,
}
\makeatletter
  \newcommandtwoopt{\citeads}[3][][]{\href{http://adsabs.harvard.edu/abs/#3}%
    {\def\hyper@linkstart##1##2{}%
     \let\hyper@linkend\@empty\citealp[#1][#2]{#3}}}
  \newcommandtwoopt{\citepads}[3][][]{\href{http://adsabs.harvard.edu/abs/#3}%
    {\def\hyper@linkstart##1##2{}%
     \let\hyper@linkend\@empty\citep[#1][#2]{#3}}}
  \newcommandtwoopt{\citetads}[3][][]{\href{http://adsabs.harvard.edu/abs/#3}%
    {\def\hyper@linkstart##1##2{}%
     \let\hyper@linkend\@empty\citet[#1][#2]{#3}}}
  \newcommandtwoopt{\citeyearads}[3][][]%
    {\href{http://adsabs.harvard.edu/abs/#3}
    {\def\hyper@linkstart##1##2{}%
     \let\hyper@linkend\@empty\citeyear[#1][#2]{#3}}}
\makeatother

\newcommand{\Msun}{M$_{\odot}$}

\definecolor{smalt(darkpowderblue)}{rgb}{0.0, 0.2, 0.6}
\definecolor{forestgreen(traditional)}{rgb}{0.0, 0.5, 0.0}

\defcitealias{RVJ}{RVJ}

\defcitealias{MD17}{MD17}

\newcommand{\cv}{CV}
\newcommand{\cvs}{CVs}

\newcommand{\bse}{BSE}

\begin{document}

   \title{The incidence of magnetic cataclysmic variables can be explained by the late appearance of white dwarf magnetic fields}

\titlerunning{Incidence of magnetic CVs}

\author{
          Matthias R. Schreiber\inst{1}
          \and
          Diogo Belloni
          \inst{1,2}
    }
  
\authorrunning{Schreiber, M. R. and Belloni, D.}

    \institute{Departamento de F\'isica, Universidad T\'ecnica Federico Santa Mar\'ia, Av. Espa\~na 1680, Valpara\'iso, Chile\\
              \email{matthias.schreiber@usm.cl}
              \and
              São Paulo State University (UNESP), School of Engineering and Sciences, Guaratinguetá, Brazil\\
             \email{diogobellonizorzi@gmail.com}}

   \date{Received September 15, 1996; accepted March 16, 1997}
 
  \abstract
  {Assuming that white dwarf (WD) magnetic fields are generated by a crystallization- and rotation-driven dynamo, the impact of the late appearance of WD magnetic fields in cataclysmic variables (\cvs) has been shown to potentially solve several long-standing problems of \cv~evolution. However, recent theoretical works show that the dynamo idea might not be viable and that the late appearance of WD magnetic fields might be an age effect rather than related to the cooling of the core of the WD. }
   {We investigated the impact of the late appearance of WD magnetic fields on \cv~evolution assuming that the fields appear at fixed WD ages. }
   {We performed \cv~population synthesis with the BSE code to determine the fractions of \cvs~that become magnetic at different evolutionary stages. These simulations were complemented with MESA tracks that take into account the transfer of spin angular momentum to the orbit which can cause a detached phase. }
   {We find that the observed fraction of magnetic \cvs~as a function of orbital period is well reproduced by our simulations, and that in many \cvs~the WD should become magnetic close to the period minimum. The detached phase generated by the transfer of spin angular momentum is longest for period bouncers.}
   {Interpreting the late appearance of strong WD magnetic fields as a simple age effect naturally explains the relative numbers of magnetic CVs in observed samples. As many period bouncers might detach for several gigayears,
   the late appearance of WD magnetic fields at a fixed age and independent of the core temperature of the WD can significantly reduce the predicted number of accreting period bouncers. 
   }

   \maketitle
%

\section{Introduction}

The standard theory of cataclysmic variable (\cv) evolution mostly ignores the impact that strong white dwarf (WD) magnetic fields 
may have on evolutionary sequences \citep[e.g.,][]{belloni+schreiber23-1}, and recent works have shown that this might be the reason for some of the disagreements between predictions and observations. 
Postulating that strong (i.e., $\geq1$\,MG) magnetic fields can be generated in rapidly rotating crystallizing WDs, 
\citet{schreiberetal21-1} proposed a scenario that explains several otherwise inexplicable observational facts. 
According to this model, WDs that crystallize before mass transfer starts generate a strong magnetic field as soon as accretion has spun up the WD. If the emerging magnetic field connects with that of the donor star, synchronization torques transfer spin angular momentum to the orbit, which causes the separation between the two stars to increase and the \cv~to convert into a detached binary. Angular momentum loss through gravitational radiation and/or reduced magnetic braking \citep{bellonietal20-1} then slowly shrinks the orbit until the donor fills its Roche-lobe and the system becomes a magnetic CV.

This new evolutionary sequence explains why virtually all detached close binaries containing a strongly magnetic WD observed to date are close to Roche-lobe filling and typically have periods between 
$3-5$\,hr. The sequence also explains the existence of the radio pulsing WD binaries AR\,Sco \citep{marshetal16-1} and SDSS\,J191213.72-441045.1 \citep{pelisolietal23-1}. 
In agreement with the proposed crystallization dynamo, 
\citet{bagnulo+landstreet21-1} found that also in single WDs strong magnetic fields appear late, that is, at WD ages of $2-3$\,Gyr, roughly when the WDs start to crystallize.  
In addition, the proposed dynamo seemed to explain a wide range of related phenomena \citep{bellonietal21-1,schreiberetal22-1,bellonietal24-2}. 

Recently, \citet{schreiberetal23-1} suggested that strong WD magnetic fields produced by a rotation- and crystallization-driven dynamo could even contribute to solving the problem of the missing period bouncers. 
The standard theory for the formation and evolution of 
\cvs~predicts that 
$\sim60-80$\% of all \cvs~should already have passed the period minimum
\citep{kolb93-1,goliasch+nelson15-1,bellonietal18-1}, while
the fraction derived from observations ranges from a few percent
\citep{palaetal20-1,inightetal23-1,inightetal23-2} to a quarter \citep{rodriguezetal25-1}. 
As the accretion rate decreases after a \cv~passes the period minimum,  the core temperature also decreases and the WD might eventually crystallize. Assuming that this leads to the generation of a strong WD magnetic field, the binary might detach and no longer be detectable as a period bouncer \citep{schreiberetal23-1}.  
The success in explaining the otherwise inexplicable observational facts listed above has made the crystallization-driven dynamo a popular idea for the generation of WD magnetic fields. 

However, more recent theoretical works on the origin of WD magnetic fields 
indicate two main reasons why the dynamo idea might not be viable. First, the energy in the convective zone of crystallizing WDs might not be sufficient to generate strong magnetic fields \citep[e.g.,][and references therein]{CastroTapia2024}. Second, the diffusion timescale for the field generated in the core to reach the surface might be too long. In other words, even if the dynamo were somehow able to produce strong fields, WD magnetic fields would be expected to appear even later than observed \citep{camisassaetal24-1}. 

As an alternative explanation for the late appearance of WD magnetic fields, \citet{camisassaetal24-1} suggested that WD magnetic fields are generated in the convective cores of main-sequence stars and diffuse to the surface during the WD cooling phase. The corresponding diffusion timescales could agree with the observed delay in the appearance of the WD magnetic fields. 
According to this scenario, for a given WD mass, the appearance of strong magnetic fields depends only on the age of a given WD and not fundamentally on the core temperature as in the crystallization dynamo idea. 

To keep our study model independent, for this work we   used observational constraints for the appearance of strong WD magnetic fields from single WDs and WDs in detached binaries. As WD masses in \cvs~cluster around $\sim0.8$\,\Msun~we used the age at which single WDs appear to be frequently strongly magnetic as the typical age for the emergence of WD magnetic fields in CVs.  
We then investigated the implications of magnetic field appearance as a simple age effect on \cvs~by performing a population synthesis and determining the ages of \cv~WDs at different orbital periods. We found that if strong magnetic fields appear $2-3$\,Gyr after the WD formed, the fraction of predicted magnetic WDs in \cvs~and their orbital periods broadly agree with the observations. In addition, many \cvs~should become magnetic \cvs~close to the period minimum. 
This means that the evolutionary sequence proposed by \citet{schreiberetal21-1} as well as the suggested idea for solving the problem of the missing period bouncers \citep{schreiberetal23-1} do not depend on the assumption that the fields are generated by the crystallization-driven dynamo, but instead remain plausible if a fossil field needs $2-3$\,Gyr to diffuse to the surface. 

\section{Population synthesis}

To investigate the impact of the late appearance of WD magnetic fields on populations of \cvs, we carried out \cv~population synthesis with the Binary Star Evolution 
(\bse) code \citep{hurleyetal00-1,hurleyetal02-1}, updated by \citet{bellonietal18-1}.
This version of the \bse~code includes state-of-the-art prescriptions for \cv~evolution, which allows us to accurately carry out population synthesis.

We adopted the initial binary population described in detail in \citet{bellonietal24-1}, corresponding to ${\approx6.15\times10^5}$ zero-age main-sequence binaries.
We picked the primary mass from the canonical \citet{Kroupa_2001} initial mass function in the range between $1$ and $8$~\Msun. 
The orbital period and mass ratio distributions were constructed following \citet{MD17}. 
These distributions are correlated, that is, the distribution of orbital periods depends on the primary mass as well as the binary fraction, while the distributions of eccentricity and mass ratio depend on the orbital period and the primary mass.

For common-envelope evolution, a low efficiency was adopted by assuming that $25$\% of the dissipated binary orbital energy is used to expel the common envelope, which is consistent with the evidence that cataclysmic variable progenitors experience strong orbital shrinkage during common-envelope evolution \citep[e.g.,][]{zorotovicetal10-1}.
The envelope-structure parameter depends on the red giant properties \citep{claeysetal14} and contributions from recombination energy were assumed negligible.

The subsequent evolution is driven by orbital angular momentum loss due to magnetic braking, consequential angular momentum loss, and gravitational radiation.
We assumed the standard prescription for gravitational radiation and for magnetic braking \citet{rappaportetal83-1}, with the normalization factors determined by \citet{zorotovicetal16-1}.
For consequential angular momentum loss, we adopted the prescription put forward by \citet{schreiberetal16-1}.

During the Galactic disk life-time, assumed to be ${\approx10}$~Gyr, we adopted a constant star formation rate. 
This implies that the birth-time distribution of the binary main-sequence stars is uniform.
The binary main-sequence stars were then evolved with the \bse~code from the birth time until the assumed Galactic age of $10$~Gyr. 

\section{Predicted white dwarf ages in CVs}

\begin{figure}
\begin{center}
\includegraphics[width=0.9\linewidth]{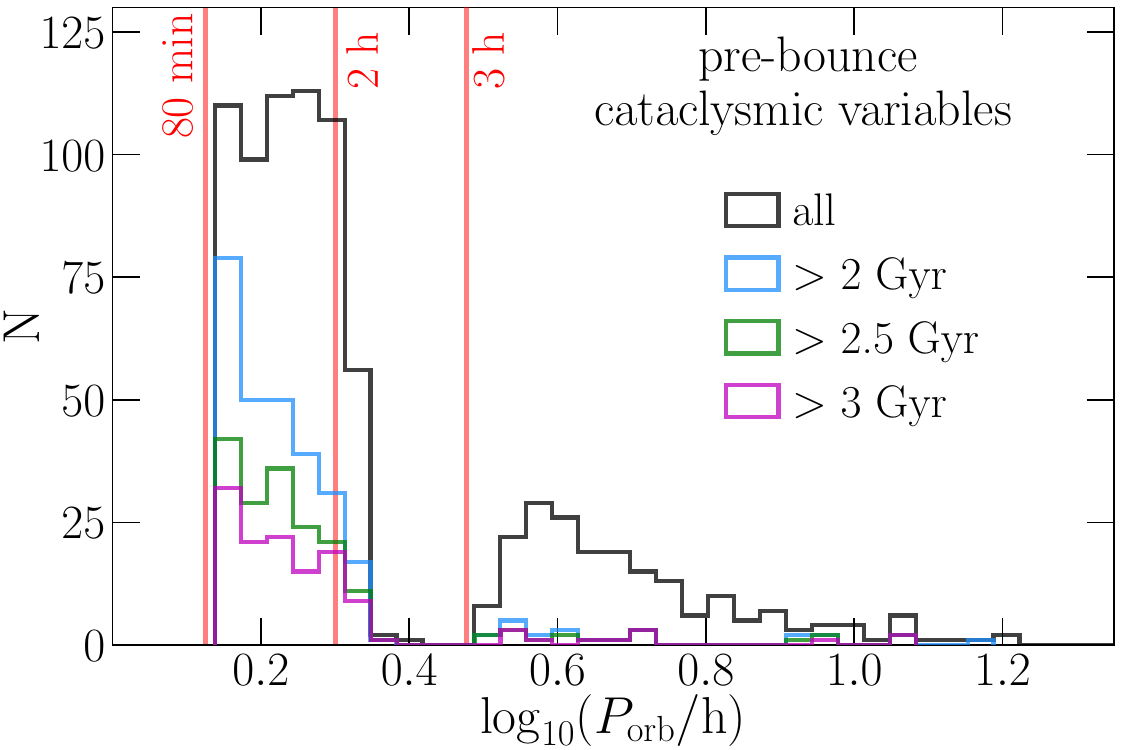}
\end{center}
\vspace{-0.5cm}
\caption{Orbital period distribution of all pre-bounce CVs. We also show the distributions of pre-bounce CVs with WDs older than $2$, $2.5$, and $3$~Gyr. The overall fractions of systems with WDs older than $2$, $2.5$, and $3$~Gyr are ${36}$\%, ${23}$\%, ${16}$\%, respectively. Above the period gap (i.e., $P_{\rm orb}\gtrsim3$~h), the same fractions are ${12}$\%, ${9}$\%, and ${6}$\%, while for systems inside and below the gap, they are ${45}$\%, ${27}$\%, and ${20}$\%.}
\label{Fig:Porb}
\end{figure}

\begin{figure}
\begin{center}
\includegraphics[width=0.9\linewidth]{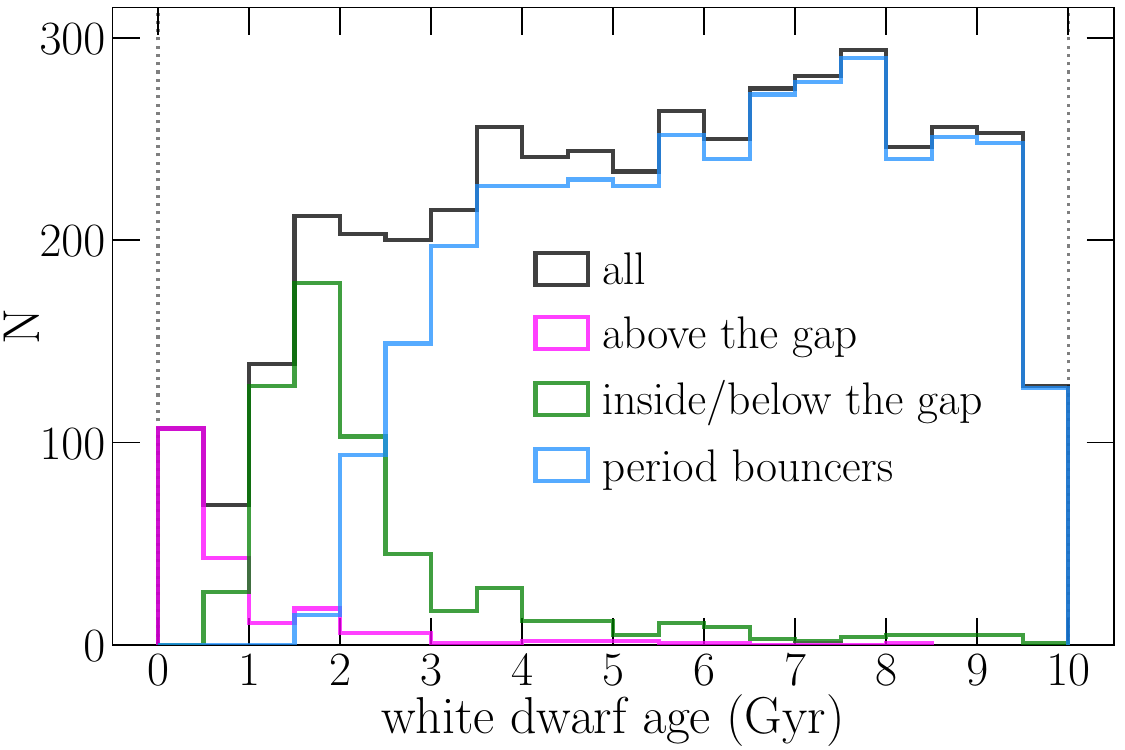}
\end{center}
\vspace{-0.5cm}
\caption{WD age distribution for different groups of \cvs. The distribution of all \cvs~is dominated by that of period bouncers at ages ${\gtrsim2-3}$~Gyr, while at younger ages it is dominated by pre-bounce systems, with peaks at ${\sim0.25}$~Gyr (above the gap) and ${\sim1.75}$~Gyr (below or inside the gap).}
\label{Fig:WD}
\end{figure}

To compare the model prediction with observations we separated the predicted \cvs~into two groups, according to the type of the donor.
If the donor is still burning hydrogen in its interior, we have a pre-bounce \cv, while if the donor is a brown dwarf, we have a period bouncer. For the threshold between the two classes we assumed $0.07$\,\Msun. We furthermore separated pre-bounce \cvs~into those with donors that still have a radiative core (with periods exceeding ${\sim3}$\,hr and donor masses ${\geq0.2}$\,\Msun) experiencing strong angular momentum loss through magnetic braking and those that have fully convective donors (periods shorter than ${\sim3}$\,hr and donor masses below $0.2$\,\Msun). 
 
To illustrate the number of strongly magnetic WDs our population synthesis predicts, we show the 
period distributions of predicted present-day \cvs~that did not yet reach the period minimum highlighting the fraction of \cvs~that are older than the assumed threshold 
for the appearance of strong WD magnetic fields (Fig.\,\ref{Fig:Porb}).  
Although most pre-bounce systems host young WDs, a significant fraction of ${36}$\%, ${23}$\%, and ${16}$\% of them host WDs older than $2$, $2.5$, and $3$~Gyr, respectively.
For systems above the period gap, that is, \cvs~with orbital periods longer than $\sim3$~h, the fractions of systems having WDs older than $2$, $2.5$, $3$~Gyr are ${12}$\%, ${9}$\%, and ${6}$\%, respectively.
The same fractions increase to ${45}$\%, ${27}$\%, and ${20}$\% for systems below or inside the gap, with orbital periods shorter than $3$~h.

In Fig.~\ref{Fig:WD} we include predicted period bouncers and show the age distributions of \cvs.
The distribution of all \cvs~is dominated by that of period bouncers at ages ${\gtrsim2}$~Gyr, while at lower ages it is dominated by pre-bounce systems, with peaks at ${\sim0.25}$~Gyr (above the gap) and ${\sim1.75}$~Gyr (below or inside the gap).
The fraction of period bouncers older than $2, 2.5,$ and $ 3$\,Gyr is 
$99.6, 96.9,$ and $ 92.8\%$, respectively. 
The absolute number of \cvs~with WDs exceeding a given age are given in Table\,\ref{Tab:Numbers}. 
\begin{figure}
\begin{center}
\includegraphics[width=0.9\linewidth]{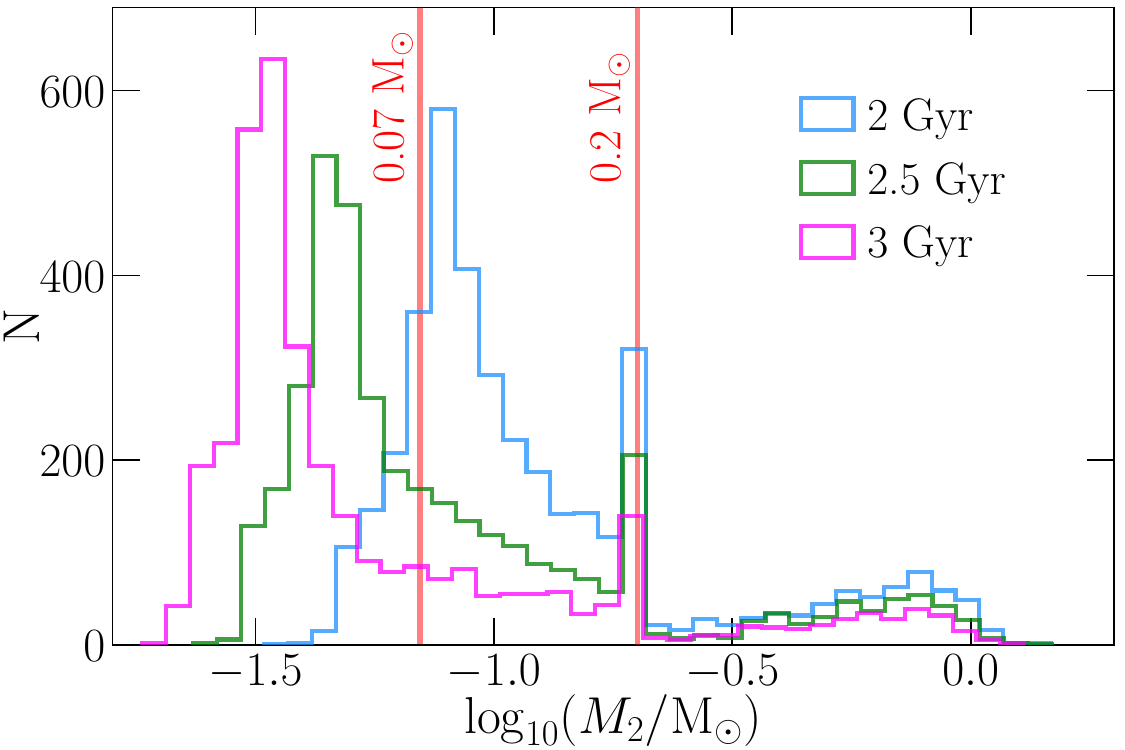}
\end{center}
\vspace{-0.5cm}
\caption{Donor mass distribution of the present-day \cvs~when their WD ages reached $2$, $2.5$, and $3$~Gyr illustrating at which donor mass a strong magnetic field could have appeared. 
The peak of the distributions is at ${0.076}$, ${0.045}$, and ${0.033}$~\Msun~for age limits of $2$, $2.5$, and $3$\,Gyr respectively. 
For ${35}$\%, ${69}$\%, and ${81}$\% of the present-day period bouncers the magnetic field appeared at donor masses $\leq0.08$\,\Msun, that is, after or around the  period minimum. 
}
\label{Fig:M2}
\end{figure}

Given that   the emergence of the magnetic field may cause a detached phase (generated by the transfer of spin angular momentum to the orbit), which might take several Gyr if the WD magnetic field appears at the period minimum or beyond (see \citet{schreiberetal23-1} and Appendix\,\ref{app}), it is important to evaluate at which period the field appeared for the predicted present-day \cvs. 
The distribution of secondary masses at which the present-day \cvs~became strongly magnetic (assuming as above three different ages for the appearance of the field) is shown in Fig.~\ref{Fig:M2}.
Interestingly, for ${\approx35}$\%, ${\approx69}$\%, and ${\approx81}$\% of  the current period bouncers the WD reached the age of  $2$, $2.5$, and $3$~Gyr when the donor mass was below $0.08$\,\Msun, that is, close to or after the system reached the period minimum. 

\begin{table}
\caption{Numbers of \cvs~predicted in our population synthesis.}
\vspace{-0.15cm}
\label{Tab:Numbers}
\centering
\setlength\tabcolsep{10pt} 
\renewcommand{\arraystretch}{1.25} 
\begin{tabular}{lrrrr}
\hline\hline
                 &    \multicolumn{4}{c}{WD age (Gyr)} \\
                 &   all       & $\geq2$     & $\geq2.5$     & $\geq3$ \\
\hline
above the gap    &    $203$    &       $24$  &         $18$  &       $12$  \\
in/below the gap &    $600$    &      $267$  &        $164$  &      $119$  \\
period bouncers  &   $3564$    &     $3549$  &       $3455$  &     $3306$  \\
total            &   $4367$    &     $3840$  &       $3637$  &     $3437$  \\
\hline
\end{tabular}
\end{table}

\section{Strongly magnetic WDs in \cvs}

We simulated \cv~populations and determined the age distribution of the accreting WDs. In what follows we estimate the number of strongly  magnetic ($\geq1$\,MG) white dwarfs in the simulated sample and compare these predictions with observations. 

\subsection{Predicting magnetic WDs in CVs}

To estimate when a strong magnetic field appears we used the available observational constraints from single WDs and WDs in detached binaries. 
First, we assumed that WDs that formed through common envelope evolution are all born without a strong magnetic field. While a few very young (i.e., $\leq0.6$\,Gyr) and massive single WDs are strongly magnetic \citep[][their fig. 2]{bagnulo+landstreet22-1}, we assumed that these WDs become strongly magnetic early through a channel that is not available to close binaries (e.g., the merging of two C/O core WDs). This is reasonable because not a single young and strongly magnetic WD has been found in thousands of detached post-common envelope binaries \citep{liebertetal05-1,schreiberetal10-1,schreiberetal21-1}. Instead, all strongly magnetic WDs found among these detached progenitors of \cvs~are older than $\sim2$\,Gyr \citep{parsonsetal21-1,vanroesteletal25-1}.

We assumed a fixed WD age for the appearance of strong WD magnetic fields of $2, 2.5,$ and $ 3$ Gyr. We selected these thresholds because of the ages of magnetic WDs in detached binaries (see above) and because the fraction of strongly magnetic (i.e., $\geq1$\,MG)  single WDs with masses typical for \cvs\, \citep[i.e., $0.75-0.95$\,\Msun\,][]{zorotovicetal11-1,palaetal22-1} significantly increases beyond these ages. In the sample analyzed by \citet[][their fig.\,2]{bagnulo+landstreet22-1} only one out of 27 ($=3.7\pm3.6\%$) WDs younger than $2$\,Gyr is strongly magnetic, while this fraction increases to $50\pm17\%$ ($4/8$) and $60\pm22\%$ ($3/5$) for WDs older than 2 and 3 Gyr, respectively.  
We thus assumed that $30-80$\,\% of white dwarfs in \cvs~become magnetic as soon as they pass the defined age threshold of $2-3$\,Gyr. The large range of possible fractions (i.e.,~$30-80$\%) of strongly magnetic WDs results from the small number of observed WDs with masses typical for \cvs~($0.75-0.95$\Msun) that are older than 2 Gyr.

Unsurprisingly, based on the above assumptions, our simulations predict a steady increase in the fraction of magnetic WDs moving from \cvs~above the gap, to \cvs~in and below the gap, to period bouncers.
Taking into account that $30-80$\% of WDs that pass the age threshold become strongly magnetic, 
the predicted fraction of strongly magnetic WDs in \cvs~is $\sim2-10$\,\% above the period gap, $6-36$\,\% in and below the gap, and reaches $28-80$\,\% for period bouncers. 
In addition, in $7-65$\,\% of all present-day period bouncers the magnetic field 
is predicted to appear when the donor star mass decreases below $0.08$\,\Msun. 
In what follows we relate these findings to observed samples of \cvs~and discuss implications for the problem of the missing period bouncers. 

\subsection{Comparison with observations}

The small but almost complete 150\,pc sample of \cvs~\citep{palaetal20-1} 
contains $1/6=17\pm14$\% magnetic systems above the gap, and this number increases for systems in and below the gap to $14/33=40\pm9$\%. 
While clearly subject to low number statistics, this is broadly consistent with the 
$2-10$ and $6-36$ \% of \cvs~with strongly magnetic WDs in our predicted \cv~population. 
Only three period bouncers are in the 150\,pc sample and none of them contains a magnetic WD. This seems to disagree with our prediction, but magnetic period bouncers might be absent in observed samples as they detach for long periods of time when the field appears close to the period minimum or beyond.  

For the larger SDSS \cv~samples from \citet{inightetal23-1,inightetal23-2,inightetal25-1} we find the fraction of magnetic \cvs~to be  
$16.5\pm2.6$\% above the period gap and
$24.4\pm2.1$\% otherwise. Both values are still in reasonable agreement with the prediction, although the predicted fraction of strongly magnetic WDs above the period gap is slightly smaller than observed (by a factor of $\sim1.5$). 
Unfortunately, it remains unclear whether the differences between the larger SDSS sample 
and the small volume-limited sample is mainly caused by observational biases or by low number statistics. However, we believe that observational biases are more likely the cause of the differences between the two samples as  the period gap also looks different for 
both samples 
\citep{schreiberetal24-1}. A large and complete  volume-limited sample of \cvs~is needed to perform more detailed comparisons between observations and theoretical predictions. The predicted and observed fractions are summarized in Table\,\ref{Tab:Mag}.

Given the general agreement, however, we conclude that if the age of the WD is the one and only essential parameter for producing strong magnetic fields, as perhaps indicated by studies of single WDs \citep{bagnulo+landstreet21-1,bagnulo+landstreet22-1}, the predicted number of magnetic \cvs~roughly agrees with the observations if most WDs in \cvs~become strongly magnetic at ages of $2-3$\,Gyr. 

\begin{table}
\caption{Observed and predicted fractions of strongly magnetic WDs in \cvs. }
\vspace{-0.15cm}
\label{Tab:Mag}
\centering
\setlength\tabcolsep{10pt} 
\renewcommand{\arraystretch}{1.25} 
\begin{tabular}{lrrr}
\hline\hline
                 &   150 pc       & SDSS     & model     \\
\hline
above gap    &    $17\pm14$\,\%    &       $16.5\pm2.6$\,\%  &         $2-10$\,\%  \\
in/below gap &    $40\pm9$\,\%   &      $24.4\pm2.1$\,\%  &        $6-36$\,\%  \\
\hline
\end{tabular}
\tablefoot{We assumed that $30-80$\,\% of 
\cvs~containing WDs older than the threshold become strongly magnetic. As the observed sample we used the $150$\,pc sample \citep{palaetal20-1} and the larger SDSS samples \citep{inightetal23-1,inightetal23-2}.}
\end{table}

\subsection{Magnetism and detachment in CVs}

As described above, an age-dependent appearance of strong magnetic fields in WDs provides a reasonable agreement between the observed and predicted fractions of magnetic WDs in \cvs. 
If this scenario for the occurrence of magnetic WDs in \cvs~is correct, a relatively large number of period bouncers should contain strongly magnetic WDs ($\sim27-80$ \%). 

As described in the introduction, \citet{schreiberetal21-1} developed an evolutionary sequence connecting detached binaries containing a magnetic WD \citep{parsonsetal21-1}, WD pulsars \citep{marshetal16-1,pelisolietal23-1}, and magnetic \cvs. 
The idea of emerging magnetic fields converting \cvs~to detached systems has later been extended to period bouncers \citep{schreiberetal23-1}. However, in the latter work, the appearance of the magnetic field was assumed to depend on the onset of crystallization, which in turn depends on the density, the composition, and the core temperature of the WD. Given the recent criticism on the crystallization dynamo idea \citep{CastroTapia2024,camisassaetal24-1} this scenario might not apply.

In Appendix \ref{app} we present MESA calculations illustrating that the same gigayear-long detachment is predicted if the magnetic field appears close to the period minimum or beyond because the WD age reached the threshold. Figure\,\ref{Fig:MESA} shows that the detached phase takes longer when the companion's mass is lower.  
From our population synthesis we estimated that in $7-65$\,\% of all present-day period bouncers, the magnetic field appeared close to the period minimum or later.
If the occurrence of the magnetic field in these systems indeed causes a gigayear-long detached phase, the predicted number of accreting period bouncers is potentially significantly reduced. 
This represents an important result because it implies that  the evolutionary sequence proposed by  \citet{schreiberetal21-1} and also the idea of reducing the predicted number of period bouncers through the appearance of WD magnetic fields do not depend on postulating the existence of a crystallization- and rotation-driven dynamo. 
It is unclear whether the crystallization- and rotation-driven dynamo can explain the relative numbers as effectively as the purely time dependent appearance of WD magnetic fields.
This is because for high accretion rates the WD core is heated on timescales of $\sim10^8$\,yr, which may prevent accreting WDs from crystallizing, and might melt previously crystallized cores \citep{schreiberetal23-1}. 

The idea of period bouncers detaching because of the appearance of the magnetic field 
predicts a large number of magnetic WDs with brown dwarf companions that are not filling their Roche-lobe. A few systems that might fit this description have been identified recently through cyclotron \citep{vanroesteletal25-1} or  X-ray \citep{munozetal23-1,cunninghametal25-1} emission. However, the interpretation of these objects as detached period bouncers remains uncertain as the observed low accretion rates might also result from low states, as first observed in EF\,Eri \citep{fioretal24-1}.

\subsection{Dependence on unconstrained parameters}

While the outlined scenario appears plausible, we have to admit that our predictions depend on a number of assumed parameters that are neither observationally nor theoretically well constrained. 
While the strength of magnetic braking in \cvs~is relatively well constrained by the measured donor star radii \citep{McAllister2019} and the onset of the period gap \citep{kniggeetal11-1,schreiberetal24-1}, magnetic braking might be significantly stronger than predicted by the standard prescription \citep{rappaportetal83-1} for binaries with longer orbital periods.

There is growing evidence that magnetic braking saturates in close binaries \citep{elbadryetal22-1,bellonietal24-1}, which implies orders of magnitude stronger angular momentum loss for detached binaries with periods between a few hours and a few days 
\citep[see also][]{barrazaetal25-1}. Stronger angular momentum loss between common envelope evolution and the second mass transfer phase (the \cv~stage) would lead to generally younger \cvs, and thus would change the fractions of $2-3$\,Gyr  \cvs~predicted by population synthesis. Assuming the same WD age for strong field appearance ($2-3$\,Gyr), fewer systems would be predicted to be magnetic among pre-bounce \cvs, while the fraction of period bouncers among all \cvs~would further increase, as would the number of period bouncers that detach.    

The second parameter that could impact the predictions is the assumed age at which the WD magnetic field appears. We simply assumed constant ages between $2$ and $3$\,Gyr when estimating the fraction of strongly magnetic WDs for different evolutionary stages (orbital period ranges). At these ages single WDs with typical CV WD masses are frequently strongly magnetic \citep{bagnulo+landstreet22-1}. However, given the small number of observed WDs in this parameter range we cannot exclude a younger age for the field appearance as, for example, the diffusion timescale for typical \cv~WD masses \citep[$0.83$\,\Msun,~][]{zorotovicetal11-1} seems to be shorter than $2$\,Gyr \citep{camisassaetal24-1}. This would have the opposite effect of decreased evolutionary timescales toward the \cv~phase, that is, this would cause the number of predicted magnetic pre-bounce \cvs~to increase and the number of detached period bouncers would decrease. The general agreement between observations and predictions would persist. 

We also note that the durations of the detached phases calculated in \citet{schreiberetal23-1} and in the Appendix assume a high efficiency for the accreted angular momentum to spin up the WD, a synchronization timescale ($1$\,Myr), and that all the spin angular momentum is transferred to the orbit. Even for magnetic fields appearing in period bouncers, the detached phase might well be  shorter than we predict here. In this case, the number of accreting period bouncers would be reduced by a smaller amount, but the prediction of a relatively large number of magnetic WDs in period bouncers would remain.  

Finally, as discussed above, the fraction of magnetic WDs in observed samples of \cvs~that have not yet passed the period minimum is slightly higher than predicted (see Table\,\ref{Tab:Mag}). 
Perfect agreement could be reached by assuming a higher fraction of WDs becoming strongly magnetic at the threshold or assuming a lower age limit for the field appearance. 
However, this would lead to predicting fewer detached period bouncers. 

We conclude that assuming that the late appearance of WD magnetic fields can be modeled as a simple age effect leads to predicted fractions of strongly magnetic \cvs~that are in general agreement but slightly lower than observed. For now, investigating the full parameter space to potentially reach even better agreement would represent a rather futile exercise as several key processes, such as angular momentum loss through magnetic braking, remain poorly understood and  representative and large samples of observed single WDs (with typical CV WD masses) and \cvs~are not yet available. 

\section{Conclusions}

We performed binary population synthesis of \cvs~and assumed that strong WD magnetic fields appear at WD ages of $2-3$\,Gyr. Our simulations predict relative numbers of magnetic \cvs~above and below the period gap that are broadly in agreement with the observations and that a large fraction of period bouncers should contain magnetic WDs. Furthermore, in $7-65$\% of the present-day \cvs~the magnetic field appeared when the system had reached or evolved past the period minimum. At this evolutionary stage the appearance of the WD magnetic fields and the transfer of spin angular momentum to the orbit through synchronization torques can cause extended  detached phases (several gigayears).  
We therefore conclude that the evolutionary sequence connecting AR\,Sco, detached binaries containing a strongly magnetic WDs \citep{parsonsetal21-1}, and magnetic \cvs~above the gap \citep{schreiberetal21-1}, and also the idea of reducing the number of predicted accreting period bouncers through detachment \citep{schreiberetal23-1}, do not depend on assuming that the fields are generated by the crystallization- and rotation-driven dynamo, but instead could be reproduced by simply assuming that strong magnetic fields on the surface of WDs appear as a result of aging.

\begin{acknowledgements}
MRS acknowledge financial support from {FONDECYT} (grant number {1221059}). DB acknowledges support from FONDECYT (grant number 3220167) and the São Paulo Research Foundation (FAPESP), Brazil, Process Numbers {\#2024/03736-2} and {\#2025/00817-4}.
\end{acknowledgements}

\bibliographystyle{aa}

\begin{appendix}
\section{MESA simulations}\label{app}

To evaluate the plausibility of the suggestion that the number of period bouncers is reduced by the late appearance of magnetic fields as suggested by \citet{schreiberetal23-1}, the duration of the detached phase generated by the appearance of magnetic fields plays an important role. The duration is of the order of several tens of Myr above the gap \citep{schreiberetal21-1} but can reach $\sim\,1$\,Gyr if coupled with the binary entering the period gap \citep[e.g.,][red track in their fig. 1]{schreiberetal23-1}. 

To evaluate how long the detached phase may take at shorter periods, we carried out MESA simulations of \cvs~that reach the age of field appearance at different orbital periods. 
We used the version r15140 of the MESA code \citep[][]{Paxton2011,Paxton2013,Paxton2015,Paxton2018,Paxton2019,Jermyn2023} to calculate \cv~evolution.
Our approach follows closely that by \citet{schreiberetal23-1}\,\footnote{\href{https://zenodo.org/records/10008722}{https://zenodo.org/records/10008722}}.
The only difference is that we assumed that the magnetic field appears on the WD surface when its age has a certain value, instead of assuming the rotation- and crystallization-driven dynamo.
To illustrate the impact of late magnetic field appearance around the period minimum and during period bouncer evolution, we assumed an age of $2$~Gyr.

We simulated four systems and started all simulations immediately after the WD formation. 
We assumed a  WD mass of $0.8$~\Msun~and an initial companion mass of $0.6$~\Msun.
The initial orbital periods ($0.4$, $0.5$, $0.6$, $0.7$~d) were chosen so that the WD magnetic fields appear close to the period minimum or during period bouncer evolution.

\begin{figure}
\begin{center}
\includegraphics[width=1\linewidth]{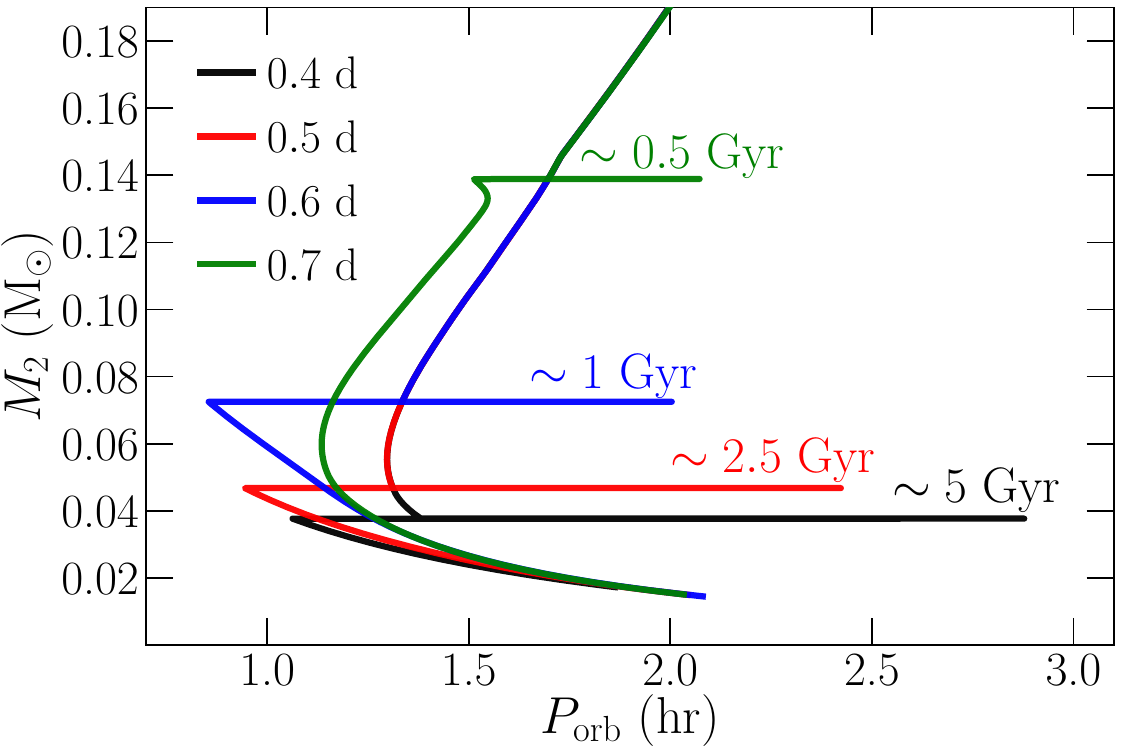}
\end{center}
\vspace{-0.5cm}
\caption{Donor mass against orbital period during evolution below the period gap. The magnetic field is assumed to appear on the WD surface when it is $2$~Gyr old and the timescales for the detached phase triggered by the magnetic field appearance are given in the figure. The lower the donor mass when the magnetic field appears, the stronger the impact of the synchronizing torque, and consequently, the longer the detachment phase. For magnetic fields appearing around the period minimum or during the period bouncer evolution, the detachment phases last ${\gtrsim1}$~Gyr.}
\label{Fig:MESA}
\end{figure}

Figure~\ref{Fig:MESA} shows the evolutionary tracks we obtained and clearly illustrates that the duration of the detached phase 
increases with decreasing donor mass at which the magnetic field appears. This is simply because the relative importance of the spin angular momentum of the WD is increasing for lower donor masses. 
The detached phase therefore can become fundamentally important for period bouncers.  

\end{appendix}

\end{document}